\documentclass{ceab}   

\usepackage{epsfig}     
\usepackage{graphicx}   

\usepackage{ceabbib}     
\usepackage[T1]{fontenc}

\begin{document}

\def\tit{Recurrent Jovian electron
increases and GCR decreases}
\def\aut{P. K\"uhl et al., 2012}

\title{Simultaneous analysis of recurrent Jovian electron
increases and galactic cosmic ray decreases}

\author{P. K\"uhl$^{1}$, N. Dresing$^{1}$, P. Dunzlaff$^{1}$, H. Fichtner$^{2}$,
J. Gieseler$^{1}$,\\ R. G\'omez-Herrero$^{1}$, B. Heber$^{1}$, A. Klassen$^{1}$,
J. Kleimann$^{2}$, A. Kopp$^{1}$,\\ M. Potgieter$^{3}$, K. Scherer$^{2}$ and
R.D. Strauss$^{3}$
\vspace{2mm}\\
\it $^1$IEAP, Christian-Albrechts-University, Kiel, Germany,\\ 
\it $^2$Theoretische Physik IV, Ruhr-University Bochum, Bochum, Germany,\\ 
\it $^3$Unit for Space Physics, North-West University, Potchefstroom,
South Africa
}

\maketitle

\begin{abstract}
The transport environment for particles in the heliosphere, e.g. galactic cosmic
rays (GCRs) and MeV electrons (including those originating from Jupiters
magnetosphere), is defined by the solar wind flow and the structure of the
embedded heliospheric magnetic field. Solar wind structures, such as
co-rotating interaction regions (CIR), can result in periodically modulation of
both particles species. A detailed
analysis of this recurrent Jovian electron events and galactic cosmic ray
decreases measured by SOHO EPHIN is presented here, showing clearly a change of
phase between both phenomena during the cause of the years 2007 and 2008. This
effect can be explained by the
change of difference in heliolongitude between the Earth and Jupiter, which is
of central importance for the propagation of Jovian electrons. Furthermore, the
data can be ordered such that
the 27-day Jovian electron variation vanishes in the sector which does not
connect the Earth with Jupiter magnetically using observed solar wind speeds.
\end{abstract}


\section{Introduction}
The sun constantly emits the solar wind, which can be further categorized in
a slow (400km/s) and a fast solar wind (700km/s). The magnetic field of the
source regions is frozen in the solar wind plasma and forms so-called Parker
spirals. If a fast wind, streaming from a coronal hole, follows a slow stream
they interact with each other forming a Stream Interaction Region (SIR). If the
coronal hole is stable over a long period these SIRs can be observed every
rotation and therefore are called Co-Rotating Interaction Regions (CIRs).
They are associated with rapid changes in the plasma signatures like solar wind
speed, density and temperature as well as in the magnetic field. As
the particle transport of galactic cosmic rays (GCR) depends on the plasma
conditions, CIRs modulate the GCR intensity. If a CIR passes, one often measures
a GCR intensity decrease. If the CIR
persists over several rotations, recurrent cosmic ray decreases (RCRDs) can be
observed. Besides GCRs, the CIR also influences the propagation of MeV
electrons in the heliosphere, including those from Jupiter's magnetosphere. The
latter is known to be a point-source of electrons with energies up to 30MeV
\citep{mcdonald,simpson}. When injected into the heliosphere, these electrons
undergo the same transport processes as GCRs. Such a recurrent variation
has been reported already by \citet{chenette}. Due to the
unique geometry a strong dependence of the electron flux with Jupiter's position
relative to the observer is expected, leading to a 13 month
variation at the Earth's orbit \citep{mcdonald}. Since the curvature of the
magnetic fieldline connecting Jupiter and the observer depends on the solar wind
speed, the particle flux should vary accordingly. \newline
In this work, the influence of CIRs on the propagation of GCRs and MeV
electrons is investigated and compared to each other using measurements
from SOHO/EPHIN \citep{ephin}, SOHO/CELIAS \citep{celias} and ACE/MAG
\citep{acemag}. In addition, the general dependency of the intensity of MeV
electrons near earth on the difference in heliolongitude between the Earth and
Jupiter is examined.

\section{Simultaneous analysis of GCR and MeV electrons}
\begin{figure}[!tbp]
\centering
\includegraphics[width=0.45\linewidth]{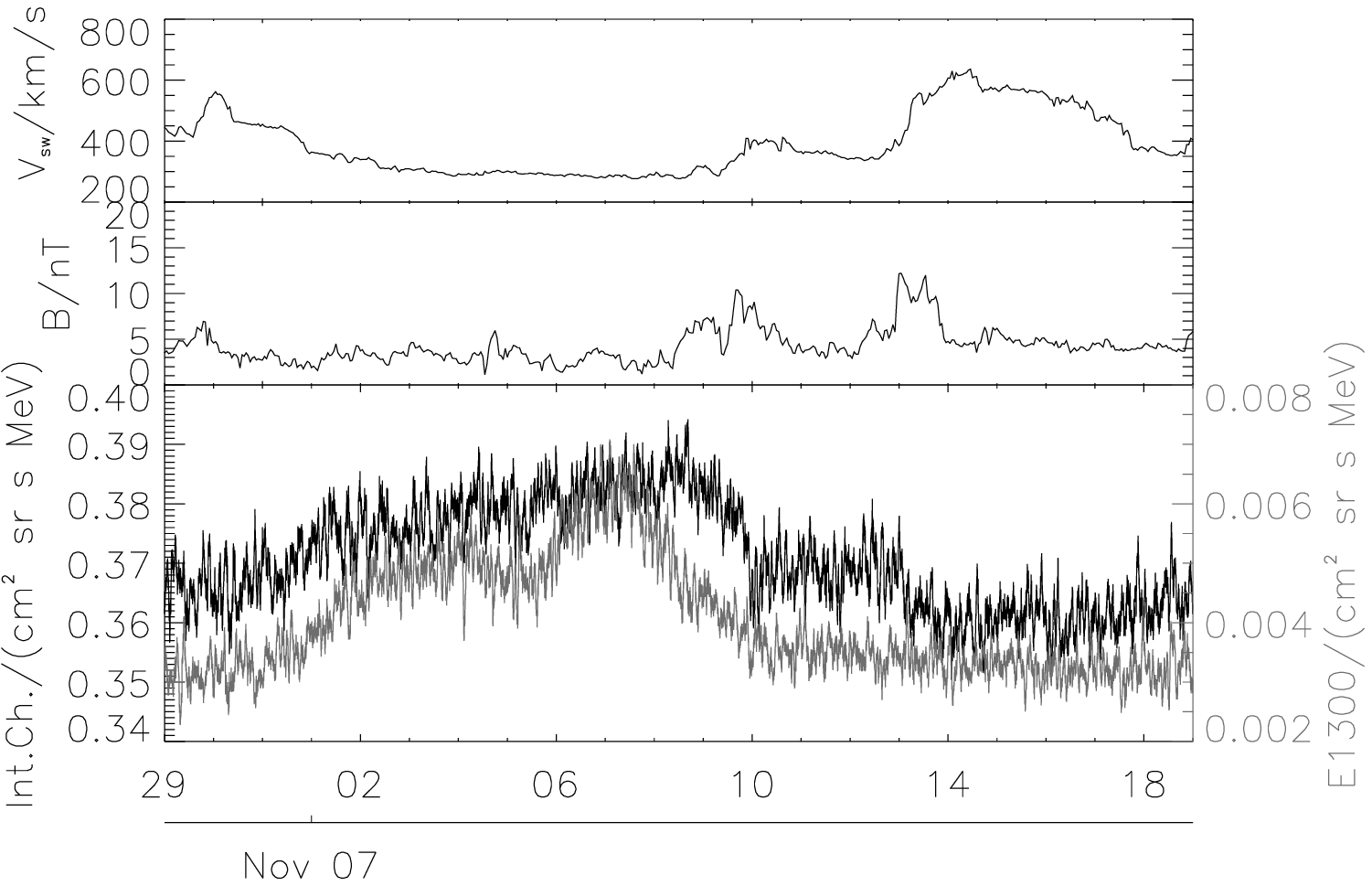}
\includegraphics[width=0.45\linewidth]{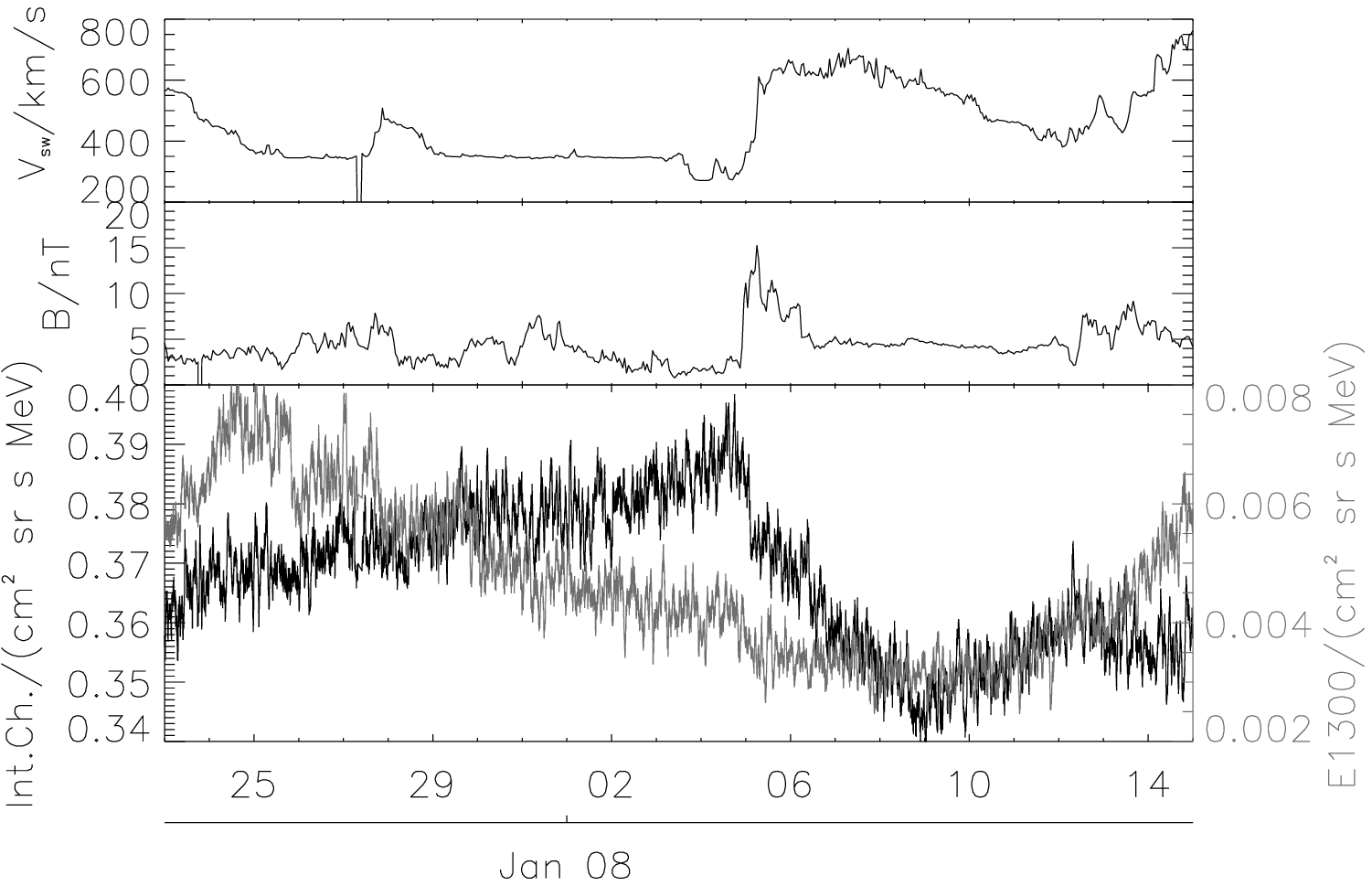}
\caption{Solar wind speed (CELIAS), magnetic field strength (ACE/MAG), GCR and
MeV electron intensities (EPHIN) in November 2007 (left) and January 2008
(right).}
\label{fig:2}
\end{figure}

To investigate the difference in modulation of GCRs and Jovian electrons, two
timeseries are shown in figure \ref{fig:2}. Results of models from \citet{Kota}
and recently \citet[]
[and references therein]{Wawrzynczak} show that convection due to the increase
in solar wind speed as well as an enhanced diffusion due to higher magnetic
field strength are responisble for the occurence of RCRDs. The
solar wind speed and the magnetic field strength are displayed in the upper two
panels of figure 1 left and right. The intensity of $>50\ MeV$ protons and
$2.6-10\ MeV$
electrons are shown in the lower panels of the same figure. As discussed in
detail by
\citet{richardson} an intensity decrease of GCRs is well correlated with the
occurrence of CIRs as indicated by the solar wind speed and magnetic field
strength increase in fig.1 left and right. In contrast, the behaviour of
the electrons is quite different. While in Nov. 2007 (fig.1, left) the
intensities of GCR and MeV electrons are decreasing simultanously (i.e. they are
in correlated), the electron intensity is decreasing several days before
the GCRs in Jan 2008 (fig.1, right). However, it is worthwhile to note that a
second decrease occurs in the electrons when the CIR passes the
spacecraft. \newline
\begin{figure}[!tbp]
\centering
\includegraphics[width=0.29\linewidth]{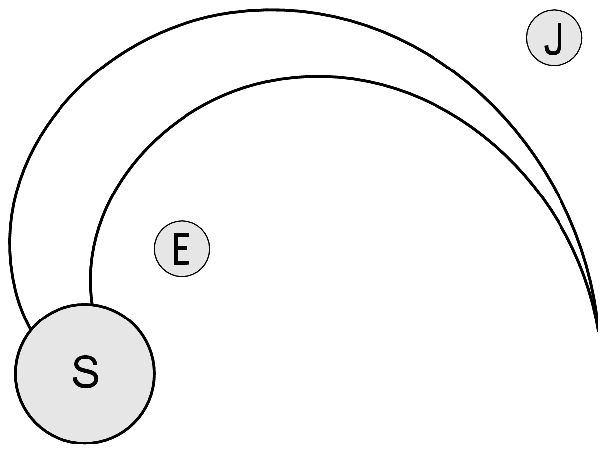}
\hspace{2.5cm}
\includegraphics[width=0.31\linewidth]{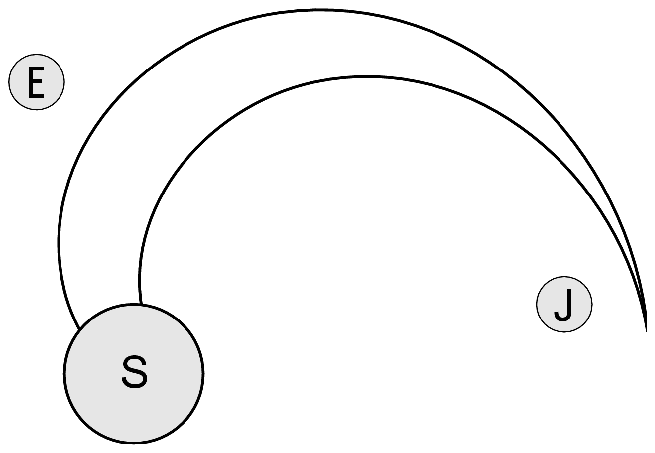}
\caption{Schemes of the CIR position relative to the Earth and Jupiter. The CIR
reaches at first the Earth (left) or Jupiter (right) respectively.}
\label{fig:3}
\end{figure}
If we assume that Jovian electrons dominate the MeV electron intensities at
$1\ AU$, the measured behaviour can be explained by the unique propagation
conditions. In the left sketch of figure 2, the CIR struture
first reaches the Earth, before crossing Jupiter. In this situation, an
observer at Earth would observe a simultaneously decrease in GCR and MeV
electrons as the CIR crosses the Earth. The right sketch shows the same CIR
structure with different positions of the Earth with respect to Jupiter, i.e.
the heliolongitude of the Earth has changed with respect to Jupiter. In this
situation, the CIR passes first Jupiter and then several days later the Earth.
Since the source of GCRs is isotropic, the relative position of the Earth to
Jupiter does not change the correlation between the occurrence of the CIR and
the GCR intensity decrease. In contrast, the MeV electron intensity
will start to decrease, when the CIR passes Jupiter. \newline
\begin{figure}[!tbp]
\centering
\includegraphics[width=1\linewidth]{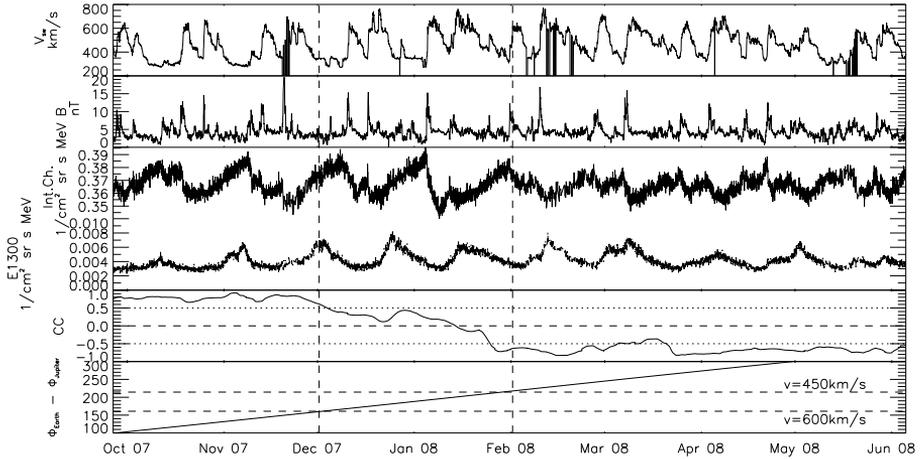}
\caption{Solar wind speed (CELIAS), magnetic field strength (ACE/MAG), GCR and
MeV electron intensities from October 2007 till June 2008. In addition, the
correlation coefficient (CC) between the intensities as well as the difference
in heliolongitude between the Earth and Jupiter is shown.}
\label{fig:4}
\end{figure}
Analysis of the orbits of Earth and Jupiter are able to match the
timeseries from figures \ref{fig:2} (left) and \ref{fig:2} (right) to the
situations shown in the figures \ref{fig:3} (left) and \ref{fig:3} (right)
respectively, reconciling the given explanation of the measured
intensities.\newline
To further analyse the effect, the plasma parameter and the
intensities
of GCR and MeV electrons are shown for the time period from Oct. 2007 to Jun.
2008) in figure \ref{fig:4}. The correlation coefficient between
the two intensities as well as the difference in heliolongitude of the Earth and
Jupiter are shown in the two lower panels. The differences in
heliolongitude, at which
parker spirals of $450$ and $600\ km/s$ magnetically connect the Earth and
Jupiter are indicated in the lowest panel by daashed lines. \newline
Assuming velocities of the CIR stream interface between $450$ and $600\ km/s$,
the difference in heliolongitude shows, that 1) from Oct. 2007 on to Dec. 2007
the CIR first crosses the Earth before reaching Jupiter, 2) between Dec. 2007
and Feb. 2008, the CIR can reach either the Earth or Jupiter first, depending on
the actual stream interface velocity and 3) from Feb. 2008 to Jun.
2008, the CIR crosses Jupiter before reaching the Earth. According to the
discussion above, this would cause the GCR and MeV electron intensities to be in
correlation before Dec. 2007 and in anticorrelation after Feb. 2008, including
a transition region inbetween. Using the correlation coefficient in the fourth
panel of fig. 3, it can be easily shown, that the behaviour of the measured
intensities is in
good agreement with the expectations discussed above.\newline
\begin{figure}[tbp]
\centering
\includegraphics[width=0.4\linewidth]{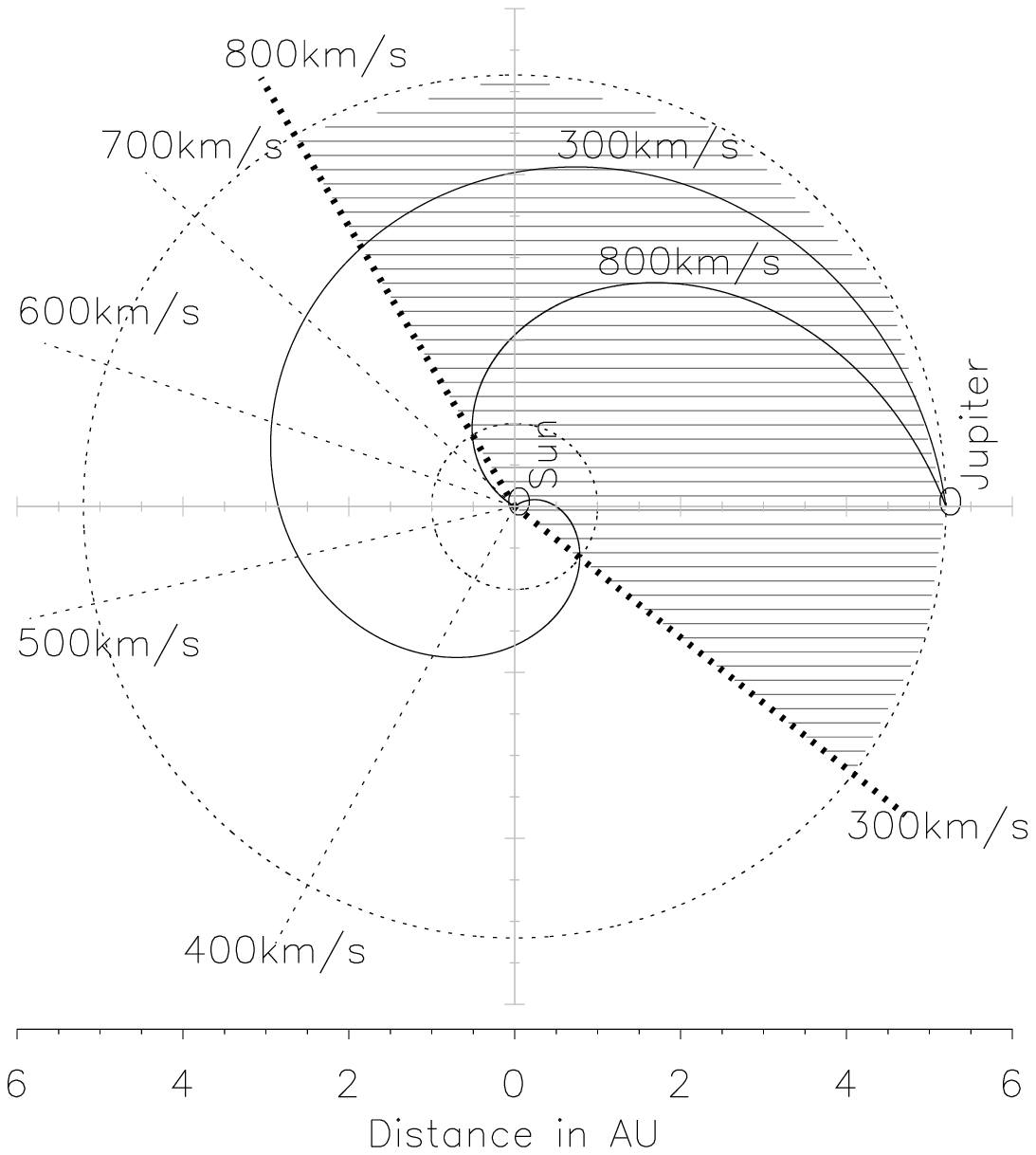}
\includegraphics[width=0.4\linewidth]{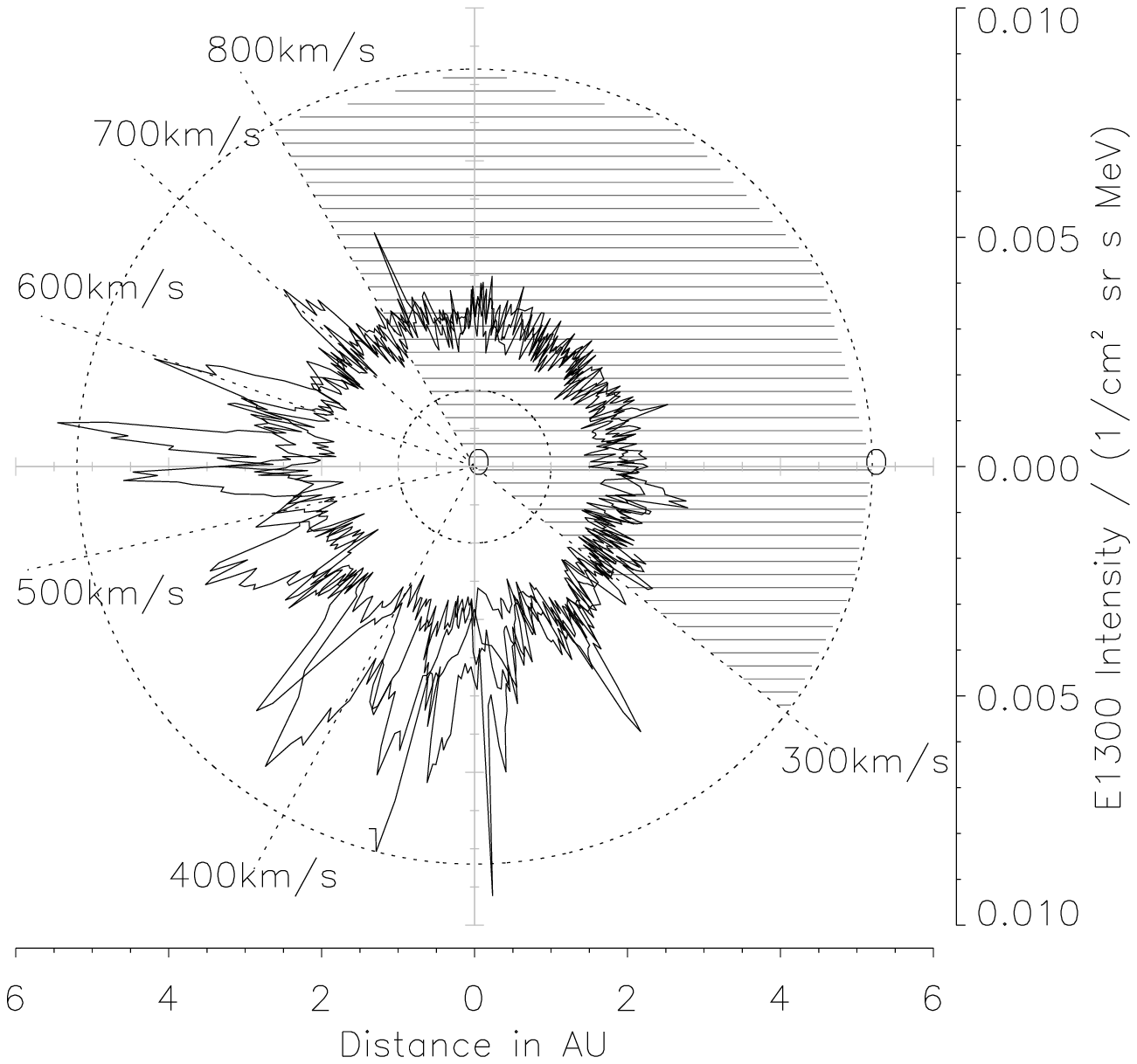}
\caption{left: coordinate system in the ecliptic with fixed position of
Jupiter. right:
same as left but with overlayed MeV electron intensity (EPHIN) dependend on the
position of the spacecraft relative to Jupiter.}
\label{fig:5}
\end{figure}
The unique propagation conditions for Jovian electrons can also be
investigated by analysing the flux of MeV electrons as function of the magnetic
connection between the observer and Jupiter. Therefore it is beneficial to
neglect any latitudinal variations and to use a coordinate system with both the
Sun and Jupiter fixed, as shown in figure \ref{fig:5} (left). Besides the Sun
(in the centre), Jupiter (on the right side) and the orbit of the Earth (dashed
circle), Parker spirals connecting the sun and Jupiter with $300\ km/s$ and
$800\ km/s$ are shown. The relative position between Earth and Jupiter can be
separated into two regions with speeds of $300\ km/s$ and $800\ km/s$ as extreme
cases. In the dashed region, magnetic connection between Earth and Jupiter can
only occur with quite unusual solar wind speeds below $300\ km/s$ or above $800\
km/s$. In the other region solar wind speeds between $300\ km/s$ and $800\ km/s$
can establish a magnetic connection. Since Jovian electrons are only able to
enter the dashed region by perpendicular transport, which is much less likely
than
the parallel one \citep{zhang}, a lower flux is expected. In Figure
\ref{fig:5} (right) the intensity of $2.64-10\ MeV$ electrons (measured with
SOHO/EPHIN) are shown as a function of the angle between SOHO and Jupiter.
In agreement with our expectations, no electron increases were detected in the
dashed region. In contrast, large intensity peaks indicating a
high jovian
electron flux increase can be found in the section, where magnetic connection
between Earth and Jupiter can be easily established with usually observed solar
wind
speeds.
\section{Summary}
In this work, the modulation of GCRs and MeV electrons by CIRs
was analysed, clearly showing a change of phase between the depressions of both
 particle populations. Investigating the difference in heliolongitude between
the Earth and Jupiter, we were able to explain this effect. Furthermore, a
region exists, which is filley by jovian electrons only via perpendicular
transport. The absence of electron increases in this region can be interpreted
as a small efficiency of perpendicular diffusion compared to the
parallel one. 

\section*{Acknowledgements} 
We thank the ACE MAG
instrument team as well as the CELIAS instrument team for providing their data.
The SOHO/EPHIN project is
supported under Grant 50 OC 0902 by the German Bundesministerium f\"ur
Wirtschaft
through the Deutsches
Zentrum f\"ur Luft- und Raumfahrt (DLR).

\bibliographystyle{ceab}
\bibliography{kuehl}

\end{document}